\documentclass{article}
\topmargin - 1.5 cm 
\begin{document}

\begin{center}
{\large{\bf Four-dimensional double singular oscillator}}
\end{center}

\begin{center}
{\bf Mara Petrosyan}
\end{center}
\begin{center}
Artsakh State University, Stepanakert
and Yerevan State University, Yerevan, Armenia
\end{center}

\begin{abstract}
The Schr\"odinger equation for the four-dimensional double
singular oscillator is separable in Eulerian, doble polar and
spheroidal coordinates in ${\rm I \!R}^4$. It is shown that
the coefficients for the expansion of double polar basis in
terms of the Eulerian basis can be expressed through the
Clebsch-Gordan coefficients of the group $SU(2)$ analytically
continued to real values of their arguments. The coefficients
for the expansions of the spheroidal basis in terms of the
Eulerian and double polar bases are proved to satisfy
three-term recursion relations.
\end{abstract}

\section{Introduction}     

The Schr\"odinger equation for a four-dimensional double singular
oscillator has the form \cite{mardoyan1}
\begin{eqnarray}
{\hat H}\psi= -\frac{\hbar^2}{2\mu}
\frac{\partial^2 \psi}{\partial u^2_i} +
\left(\frac{\mu\omega^2u^2}{2}+\frac{c_1}{u_0^2+u_1^2} +
\frac{c_2}{u_2^2+u_3^2}\right)\psi=E\psi, \label{schr-so}
\end{eqnarray}
where $u_i \,\,(i=0,1,2,3)$ are Cartesian coordinates of the space
${\rm I \!R}^4$ and $c_{1}$ and $c_{2}$ nonnegative constants.

In our paper \cite{mardoyan1} it is shown that the four-dimensional
double singular oscillator and generakized MIC-Kepler problem decribed
by the Schr\"odinger equation \cite{mardoyan2}
\begin{eqnarray}
\frac{1}{2\mu}\left(-i{\bf{\nabla}} \frac{e}{c}s{\bf A}\right)^2\,\psi
+\left[\frac{\hbar^2s^2}{2\mu r^2}-\frac{2}{r}+\frac{\lambda_1}{r(r+z)} +
\frac{\lambda_2}{r(r-z)}\right]\psi=\epsilon\psi, \label{schr-gmic}
\end{eqnarray}
are dual. Here the vector potential ${\bf A}$ corresponds to the Dirac
monopole with the magnetic charge $g=\hbar cs/e$
$\left(s=0,\pm1/2,\pm 1,\ldots\right)$ and has the form
\begin{eqnarray*}
{\bf A}=\frac{1}{r(r-z)}(y,-x,0), \qquad {\rm and} \qquad {\rm
rot}{\bf A}=\frac{{\bf r}}{r^3}.
\end{eqnarray*}
The transformation of duality is the generalization version of the
so-called Kustaanheimo-Stiefel transformation
\begin{eqnarray}
x+iy &=& 2\left(u_0+iu_1\right)\left(u_2+iu_3\right), \nonumber \\ [2mm]
z &=& u_0^2+u_1^2-u_2^2-u_3^2 \label{gks}\\ [2mm]
\gamma &=& \frac{i}{2}\ln\frac{\left(u_0-iu_1\right)\left(u_2+iu_3\right)}
{\left(u_0+iu_1\right)\left(u_2-iu_3\right)} \nonumber
\end{eqnarray}
supplemented with the ansatz $s \to -i{\partial}/{\partial \gamma}$.
We also noted, that the parameters of these systems are connected
with each other by the relations:
\begin{eqnarray*}
E=4e^2, \qquad \epsilon = -\frac{\mu \omega^2}{8}, \qquad c_a=2\lambda_a,
\qquad {\rm where} \qquad a=1,2.
\end{eqnarray*}
The first two lines of (\ref{gks}) are the transformation
${\rm I \!R}^4 \to {\rm I \!R}^3$ suggested by Kustaanheimo and Stiefel
for regularization of the equations of the celestial mechanics
\cite{ks}. Later, this transformation found other applications as well
\cite{barut,kibler}. The generalized Kustaanheimo-Stiefel transformation
(\ref{gks}) (the Kustaanheimo-Stiefel transformation supplemented with
the angle $\gamma$) was used for "synthesis" of the charge-dyon
system from the four-dimensional isotropic oscillator \cite{nerter}.

It should be noted that equation (\ref{schr-gmic}) for
$\lambda_a =0$ and $s \neq 0$ reduces tj the Schr\"odinger
equation for the MIC-Kepler system \cite{zw,mic}.
At $s=0$, the Schr\"odinger equation (\ref{schr-gmic}) is reduced to
the Schr\"{o}dinger equation for the generalized Kepler-Coulomb system
\cite{KMP}. In the case
when $s=0$ and $\lambda_1=\lambda_2\neq 0$, Eq.~(\ref{schr-gmic}) reduces
to the system suggested by Hartmann, at was used for explanation
of the spectrum of the benzene molecule \cite{H1,H2,H3}.

\section{Eulerian and Double Polar Bases}

Determine the Eulerian coordinates in ${\rm I \!R}^4$ as follows:
\begin{eqnarray}
u_0+iu_1 = u\,\cos\frac{\beta}{2}\,e^{i\frac{\alpha+\gamma}{2}},
\qquad
u_2+iu_3 = u\,\sin\frac{\beta}{2}\,e^{i\frac{\alpha-\gamma}{2}},
\label{ecoo}
\end{eqnarray}
where $u \in [0, \infty)$, $\alpha \in [0, 2\pi)$, $\beta \in [0, \pi]$,
$\gamma \in [0, 4\pi)$. In these coordinates the differential elements
of length and volume, and the Laplace operator have the form
\begin{eqnarray*}
dl^2 &=& du^2 + \frac{u^2}{4}\,\left(d\alpha^2 + d\beta^2 +
d\gamma^2+2\cos\beta\,d\alpha\,d\gamma\right), \\ [2mm]
dV &=& \frac{u^3}{8}\,\sin\beta \, du\,d\alpha \,
d\beta \, d\gamma, \\ [2mm]
\frac{\partial^2}{\partial u_i^2} &=& \frac{1}{u^3}
\frac{\partial}{\partial u}\,\left(u^3\,
\frac{\partial}{\partial u}\,\right) -\frac{4}{u^2}\,{\hat J}^2,
\end{eqnarray*}
where
\begin{eqnarray*}
{\hat J}^2 = -\left[\frac{1}{\sin\beta}\,
\frac{\partial}{\partial\beta}\left(\sin\beta\,
\frac{\partial}{\partial\beta}\right)+\frac{1}{\sin^2\beta}\,
\left(\,\frac{\partial^2}{\partial\alpha^2} -2\cos\beta\,
\frac{\partial^2}{\partial\alpha \partial\gamma}+
\frac{\partial^2}{\partial\gamma^2}\right)\right]
\end{eqnarray*}
is the square of the angular momentum operator.

The Schr\"odinger equation (\ref{schr-so}) in the Eulerian coordinates
(\ref{ecoo}) may be solved by seeking a wave function $\psi$ of the form
\begin{eqnarray}
\psi \left(u,\alpha,\beta,\gamma\right)=
R\left(u\right)\,Z\left(\alpha,\beta,\gamma\right).
\label{wv-eu}
\end{eqnarray}
This amouns to finding the eigenfunctions of
$\left\{{\hat H},\,{\hat \Lambda},\,{\hat J}_3,\,{\hat J'}_3\right\}$
of commuting operators, where the constant of motion
${\hat \Lambda}$ reads
\begin{eqnarray}
{\hat \Lambda}={\hat J}^2+\frac{\mu}{\hbar^2}\left(
\frac{c_1}{1+\cos\beta}+\frac{c_2}{1-\cos\beta}\right)
\label{Lambda1}
\end{eqnarray}
and which in the Cartesian coordinates $u_i$ has the form
\begin{eqnarray}
{\hat \Lambda}=-\frac{1}{4}\left(u^2\frac{\partial^2}{\partial u_i^2}-
u_iu_j\frac{\partial^2}{\partial u_i\partial u_j} -
3u_i\frac{\partial}{\partial u_i}\right)+\frac{\mu u^2}{2\hbar^2}
\left(\frac{c_1}{u_0^2+u_1^2}+\frac{c_2}{u_2^2+u_3^2}\right).
\label{Lambda2}
\end{eqnarray}
The operators ${\hat J}_3$ and ${\hat J'}_3$ are defined as follows:
\begin{eqnarray*}
{\hat J}_3\,\psi = -\frac{\partial\psi}{\partial\alpha} = m\psi,
\qquad
{\hat J'}_3\,\psi = -\frac{\partial\psi}{\partial\gamma} = s\psi.
\end{eqnarray*}

After substitution the expression (\ref{wv-eu}) into the
Eq.~(\ref{schr-so}) the variables in the
Schr\"odinger equation (\ref{schr-so}) are separated and we arrive at the
following system of coupled differential equations:
\begin{eqnarray}
\frac{1}{\sin\beta}\frac{\partial}{\partial\beta}\,\left(\sin\beta\,
\frac{\partial Z}{\partial\beta}\,\right) +
\frac{1}{2\left(1+\cos\beta\,\right)}\,\left[
\left(\frac{\partial}{\partial\alpha}\,+
\,\frac{\partial}{\partial\gamma}\,\right)^2-\frac{2\mu c_1}{\hbar^2}
\right]\,Z + \nonumber \\
\label{angle-eq} \\
+ \frac{1}{2\left(1-\cos\beta\,\right)}\,\left[
\left(\frac{\partial}{\partial\alpha}\,-
\,\frac{\partial}{\partial\gamma}\,\right)^2-\frac{2\mu c_2}{\hbar^2}
\right]\,Z = -\lambda Z, \nonumber    \\ [3mm]
\frac{1}{u^3}\frac{d}{d u}\,\left(u^3\,\frac{dR}{d u}\,\right)-
\frac{4\lambda}{u^2}\,R +
\frac{2\mu}{\hbar^2}\,\left(E-\frac{\mu\omega^2 u^2}{2}\right)\,R=0,
\label{rad-eq}
\end{eqnarray}
where $\lambda$ is the separation constant which is the eigenvalue
of the operator (\ref{Lambda1}).

The solution of Eq.~(\ref{angle-eq}) is easily found to be
\begin{eqnarray}
Z_{jms}\left(\alpha, \beta, \gamma; \delta_1, \delta_2\right)=
N_{jms}\left(\delta_{1},\delta_{2}\right)
\left(\cos\frac{\beta}{2}\right)^{m_1}
\left(\sin\frac{\beta}{2}\right)^{m_2}
P_{j-m_+}^{(m_2,m_1)}(\cos\beta) e^{im\alpha}\,e^{is\gamma},
\label{fZ}
\end{eqnarray}
where
$m_{1,2}=|m \pm s|+\delta_{1,2}=\sqrt{(m \pm s)^2+2\mu c_{1,2}/\hbar^2}$,
$m_{+}=(|m+s|+|m-s|)/2$ and $P_{n}^{(a,b)}(x)$ denotes a Jacobi polynomial.
The quantum number $j$ characterizes the total angular momentum and for
the (half)integer $j$ the quantum numbers $m$ and $s$ are (half)integer.
At a fixed value $j$ the $m$ and $s$ run through values:
$m,s=-j,-j+1,\ldots,j-1,j$.

Farthemore, the separation constant $\lambda$ is quantized as
\begin{eqnarray}
\lambda = \left(j+\frac{\delta_{1}+\delta_{2}}{2}\right)
\left(j+\frac{\delta_{1}+\delta_{2}}{2}+1\right). \label{lambda}
\end{eqnarray}
The normalization constant $N_{jm}(\delta_{1}, \delta_{2})$ in
(\ref{fZ}) is given (up to a phase factor) by
\begin{eqnarray*}
N_{jms}(\delta_1, \delta_2)=(-1)^{\frac{m-s+|m-s|}{2}}\,
\sqrt{\frac{(2j+\delta_1+\delta_2+2)(j-m_+)!
\Gamma(j+m_+ +\delta_1+\delta_2+1)}{16\pi^2\,\Gamma(j+m_-
+\delta_1+1) \Gamma(j-m_- +\delta_{2}+1)}},
\end{eqnarray*}
where $m_{-}=(|m+s|-|m-s|)/2$ and we assume that
\begin{eqnarray*}
\frac{1}{8}\,\int\limits_{0}^{\pi}\,\int\limits_{0}^{2\pi}\,
\int\limits_{0}^{4\pi}\,\sin\beta\,
Z_{j'm's'}^{*}\left(\alpha,\beta,\gamma;\delta_{1},\delta_{2}\right)
Z_{jms}\left(\alpha,\beta,\gamma;\delta_{1},\delta_{2}\right)
d\alpha\,d\beta\,d\gamma = \delta_{jj'}\delta_{mm'}\delta_{ss'}.
\end{eqnarray*}

Let us go now to the radial equation (\ref{rad-eq}). The introduction of
(\ref{lambda}) into  (\ref{rad-eq}) leads to
\begin{eqnarray*}
\frac{1}{u^3}\frac{d}{d u}\,\left(u^3\,\frac{dR}{d u}\,\right)-
\frac{1}{u^2}\,
\left(2j+\delta_1+\delta_2\right)\left(2j+\delta_1+\delta_2+2\right)R +
\frac{2\mu}{\hbar^2}\,\left(E-\frac{\mu\omega^2 u^2}{2}\right)\,R=0.
\end{eqnarray*}
The solution of this equation normalized by the condition
\begin{eqnarray}
\int\limits_{0}^{\infty}\,u^{3}\,R_{Nj}\left(u;\delta_1,\delta_2\right)
R_{N'j}\left(u;\delta_1,\delta_2\right) du = \delta_{NN'}
\label{norm1}
\end{eqnarray}
has the form
\begin{eqnarray}
R_{Nj}(u;\delta_1,\delta_2)=C_{Nj}(\delta_1,\delta_2)
\left(au\right)^{2j+\delta_1+ \delta_2}
e^{-\frac{a^2u^2}{2}}
F\left(-\frac{N}{2}+j; 2j+\delta_1+ \delta_2+2; a^2u^2\right),
\label{rad}
\end{eqnarray}
where $F(a; c; x)$ is the confluent hypergeometric function,
$a=\sqrt{\mu\omega /\hbar}$ and
\begin{eqnarray*}
C_{Nj}(\delta_1,\delta_2)=
\frac{4a^2}{\Gamma\left(2j+\delta_1+\delta_2+2\right)}
\sqrt{\frac{\Gamma\left(\frac{N}{2}+j+\delta_1+\delta_2+2\right)}
{(\frac{N}{2}-j)!}}
\end{eqnarray*}
The energy spectrum has the form
\begin{eqnarray}
E_N = \hbar\omega\,\left(N+\delta_1+ \delta_2+2\right)
\label{energy}
\end{eqnarray}
where $N$ is the principle quantum number, $N=0,1,2,\ldots$, and the
quantum number $j$ run through the values: $j=m_{+},m_{+}+1,\ldots,N/2$.

Thus, the Eulerian basis (\ref{wv-eu}) is the eigenfunction of the
operator (\ref{Lambda1}) and
\begin{eqnarray}
{\hat \Lambda}\,\psi_{Njms}=
\left(j+\frac{\delta_1+\delta_2}{2}\right)
\left(j+\frac{\delta_1+\delta_2}{2}+1\right)\psi_{Njms}.
\label{Lambda3}
\end{eqnarray}

In the limiting case $\delta_1=\delta_2=0$ we recover the familiar
results for the four-dimensional isotropic oscillator \cite{book}.

Let us consider the four-dimensional double singular oscillator in the
double polar coordinates \cite{book}. In the double polar coordinates
$\rho_1,\rho_2 \in [0, \infty)$, $\varphi_1,\varphi_2 \in [0, 2\pi)$,
defined by the formulae
\begin{eqnarray}
u_0+iu_1 = \rho_1\,e^{i\varphi_1}, \qquad
u_2+iu_3 = \rho_2\,e^{i\varphi_2},
\label{dpcoo}
\end{eqnarray}
the differential elements of length and volume read
\begin{eqnarray*}
dl^2 = d\rho_1^2 + d\rho_2^2 +\rho_1^2\,d\varphi_1^2 +
\rho_2^2\,d\varphi_2^2, \qquad
dV=\rho_1\,\rho_2\,d\varphi_1\,d\varphi_2,
\end{eqnarray*}
while the Laplace operator looks like
\begin{eqnarray*}
\frac{\partial^2}{\partial u_i^2} =
\frac{1}{\rho_1}\,\frac{\partial}{\partial \rho_1}
\left(\rho_1\frac{\partial}{\partial \rho_1}\right)
+ \frac{1}{\rho_2}\,\frac{\partial}{\partial \rho_2}
\left(\rho_2\frac{\partial}{\partial \rho_2}\right)+
\frac{1}{\rho_1^2}\,\frac{\partial^2}{\partial \varphi_1^2}
+\frac{1}{\rho_2^2}\,\frac{\partial^2}{\partial \varphi_2^2}.
\end{eqnarray*}
The substitution
\begin{eqnarray*}
\psi(\rho_1,\rho_2,\varphi_1,\varphi_2) = \frac{1}{2\pi}\Phi_1(\rho_1)
\Phi_2(\rho_2)\,e^{iM_1 \varphi_1}\,e^{iM_2 \varphi_2}
\end{eqnarray*}
where $M_1,M_2=0,\pm 1, \pm 2, \ldots$,
separates the variables in the Schr\"{o}dinger equation (\ref{schr-so})
and we arrive at the following system of equations:
\begin{eqnarray}
\frac{1}{\rho_1}\frac{d}{d \rho_1}
\left(\rho_1 \frac{d\Phi_1}{d \rho_1}\right) -
\frac{\left(\left|M_1\right|+\Delta_1\right)^2}{\rho_1^2}+
\frac{2\mu}{\hbar^2}\left(\frac{E}{2}+\frac{\hbar^2}{4\mu}\Omega -
\frac{\mu \omega \rho_1^2}{2}\right)\Phi_1 &=& 0,
\label{eq1}\\ [3mm]
\frac{1}{\rho_2}\frac{d}{d \rho_2}
\left(\rho_2 \frac{d\Phi_2}{d \rho_2}\right) -
\frac{\left(\left|M_2\right|+\Delta_2\right)^2}{\rho_2^2}+
\frac{2\mu}{\hbar^2}\left(\frac{E}{2}-\frac{\hbar^2}{4\mu}\Omega -
\frac{\mu\omega\rho_2^2}{2}\right)\Phi_2 &=& 0,
\label{eq2}
\end{eqnarray}
where $\beta$ -- is the separation constant and
$\Delta_a=\sqrt{M_a^2+2\mu c_a/\hbar^2}-\left|M_a\right|$.

These equations are analogous with the equations of the circular
oscillator in the polar coordinates \cite{flugge}. Thus, we get
\begin{eqnarray}
\psi_{N_1N_2M_1M_2}(\rho_1,\rho_2,\varphi_1,\varphi_2;\delta_1,\delta_2) =
\frac{1}{2\pi}\Phi_{N_1M_1}(a^2\rho_1^2;\Delta_1)\,
\Phi_{N_2M_2}(a^2\rho_2^2;\Delta_2)\,e^{iM_1\varphi_1}\,
e^{iM_2\varphi_2},
\label{dpwave1}
\end{eqnarray}
where
\begin{eqnarray*}
\Phi_{N_aM_a}(x) =
\sqrt{\frac{2\Gamma(N_a+\left|M_a\right|+\Delta_a+1)}{(N_a)!}}\,\,
\frac{ae^{-\frac{x}{2}}\,\, x^{\left(\left|M_a\right|+\Delta_a\right)/2}}
{\Gamma\left(\left|M_a\right|+\Delta_a+1\right)} \,\,
F(-N_a;\left|M_a\right|+\Delta_a +1; x).
\end{eqnarray*}
Here $N_1$ and $N_2$ are nonnegative integers
\begin{eqnarray*}
N_1 = -\frac{\left|M_1\right|+\Delta_1+1}{2}+\frac{E}{4\hbar\omega}+
\frac{\Omega}{8a^2}, \qquad
N_2 = -\frac{\left|M_2\right|+\Delta_2+1}{2}+\frac{E}{4\hbar\omega}-
\frac{\Omega}{8a^2}.
\end{eqnarray*}
From the last relations, taking into account (\ref{energy}), we
get that the double polar quantum numbers $N_1$ and $N_2$ are
connected with the principal quantum number $N$ as follows:
\begin{eqnarray*}
N= 2N_1 + 2N_2 + \left|M_1\right|+\left|M_2\right|.
\end{eqnarray*}
Excluding the energy $E$ from Eqs. (\ref{eq1}) and (\ref{eq2}), we
obtain the additional integral of motion
\begin{eqnarray}
{\hat \Omega} &=& \left[
\frac{1}{\rho_2}\frac{\partial}{\partial \rho_2}
\left(\rho_2\,\frac{\partial}{\partial \rho_2}\right)-
\frac{1}{\rho_1}\frac{\partial}{\partial \rho_1}
\left(\rho_1\,\frac{\partial}{\partial \rho_1}\right)+
\frac{1}{\rho_2^2}\frac{\partial^2}{\partial \varphi_2^2}-
\frac{1}{\rho_1^2}\frac{\partial^2}{\partial \varphi_1^2}\right]+
\nonumber \\
\label{Omega1} \\
&+& \frac{\mu^2\omega^2}{\hbar^2}\left(\rho_1^2-\rho_2^2\right)+
\frac{2\mu}{\hbar^2}\left(\frac{c_1}{\rho_1^2}-
\frac{c_2}{\rho_2^2}\right) \nonumber
\end{eqnarray}
with the eigenvalues
\begin{eqnarray*}
\Omega = \frac{2\mu\omega}{\hbar}
\left(2N_{1}-2N_{2}-\left|M_1\right|+\left|M_2\right|-
\Delta_{1}+\delta_{2}\right)
\end{eqnarray*}
and eigenfunctions
$\psi_{N_1N_2M_1M_2}(\rho_1,\rho_2,\varphi_1,\varphi_2;\Delta_1,\Delta_2)$,
i.e.
\begin{eqnarray}
{\hat \Omega}
\psi_{N_1N_2M_1M_2}(\rho_1,\rho_2,\varphi_1,\varphi_2;\Delta_1,\Delta_2)=
\Omega
\psi_{N_1N_2M_1M_2}(\rho_1,\rho_2,\varphi_1,\varphi_2;\Delta_1,\Delta_2).
\label{spectr2}
\end{eqnarray}
In the Cartesian coordinates, the operator ${\hat \Omega}$ can be
rewritten as
\begin{eqnarray}
{\hat \Omega} &=& \left(\frac{\partial^2}{\partial u_2^2}+
\frac{\partial^2}{\partial u_3^2}-
\frac{\partial^2}{\partial u_0^2}-\frac{\partial^2}{\partial u_1^2}\right) +
\nonumber \\
\label{Omega2} \\
&+&\frac{\mu^2\omega^2}{\hbar^2}\left(u_0^2+u_1^2-u_2^2-u_3^2\right)+
\frac{2\mu}{\hbar^2}\left(\frac{c_1}{u_0^2+u_1^2}-
\frac{c_2}{u_2^2+u_3^2}\right). \nonumber
\end{eqnarray}

\section{Connection Between Eulerian and Double Polar Bases}

At fixed energy values we write down the double polar bound states
(\ref{dpwave1}) as a coherent quantum mixture of Eulerian bound states
\begin{eqnarray}
\psi_{N_1N_2M_1M_2} =
\sum_{j,m,s}\,W^{jms}_{N_{1}N_{2}M_1M_2}\left(\delta_{1},
\delta_{2}\right)\,\psi_{Njms}.
\label{inter}
\end{eqnarray}
Our goal is the derivation of an explicit form of the coefficient
$W^{jms}_{N_{1}N_{2}M_1M_2}$. First, we should like to note that
from the comparison of (\ref{ecoo}) with (\ref{dpcoo}) we have
\begin{eqnarray}
\rho_1 = u\cos\frac{\beta}{2}, \qquad
\rho_2 = u\sin\frac{\beta}{2}, \qquad
\varphi_1 =\frac{\alpha+\gamma}{2}, \qquad
\varphi_2 =\frac{\alpha-\gamma}{2}.
\label{conect}
\end{eqnarray}
In relation (\ref{inter}), according to (\ref{conect}), we pass from
the double polar coordinates to the Eulerian ones. Then, substituting
$\beta=0$, taking account of
\begin{eqnarray*}
P_{n}^{(a, b)}(1) = \frac{(a+1)_n}{n!},
\end{eqnarray*}
and using the orthogonality condition for radial wave functions in
hypermomentum \cite{book}
\begin{eqnarray*}
\int\limits_{0}^{\infty}\,
R_{Nj'}\left(u;\delta_{1},\delta_{2}\right)
R_{Nj}\left(u;\delta_{1},\delta_{2}\right)\,u du =
\frac{2a^2\,\delta_{jj'}}
{\left(2j+\delta_{1}+\delta_{2}+2\right)},
\end{eqnarray*}
we obtain the following integral representation for the coefficients
$W^{jms}_{N_{1}N_{2}M_1M_2}$:
\begin{eqnarray}
W^{jms}_{N_{1}N_{2}M_1M_2}\left(\delta_{1}, \delta_{2}\right) =
\frac{\sqrt{\left(2j+\delta_{1}+\delta_{2}+1\right)(j-m_+)!}}
{\Gamma(m_2+1)\Gamma(2j+\delta_{1}+\delta_{2}+2)}\,E_{N_1N_2}^{jms}\,
K^{NN_1}_{jms}\,\delta_{M_1,m+s}\,\delta_{M_2,m-s}.
\label{inter1}
\end{eqnarray}
Here
\begin{eqnarray}
E_{N_1N_2}^{jms} &=&
\sqrt{\Gamma\left(\frac{N}{2}+j+\delta_{1}+\delta_{2}+2\right)} \times
\nonumber \\
\label{E} \\
&\times&\left[\frac{\Gamma\left(j-m_- +\delta_{1}+1\right)
\Gamma(N_1+m_1+1)\Gamma(N_2+m_2+1)}
{(N_1)!(N_2)!(\frac{N}{2}-j)!\Gamma\left(j+m_-
+\delta_{2}+1\right)\Gamma\left(j+m_+
+\delta_{1}+\delta_{2}+1\right)}\right]^{\frac{1}{2}}, \nonumber
\end{eqnarray}
and
\begin{eqnarray}
K^{Nn_1}_{jms} = \int\limits_{0}^{\infty}
e^{-x}x^{j+m_1+m_2+\delta_{1}+\delta_{2}}F\left(-N_1; m_1+1;x\right)
F\left(-\frac{N}{2}+j; 2j+\delta_{1}+\delta_{2}+2; x\right)dx,
\label{K}
\end{eqnarray}
where $x=a^2u^2$. Further, in (\ref{K}) writing down the confluent
hypergeometric function $F\left(-N_1; m_1+1;x\right)$ as a polynomial,
integrating by the formula \cite{landau}
\begin{eqnarray*}
\int\limits_{0}^{\infty}
e^{-\lambda x}x^{\nu}F\left(\alpha; \gamma; kx\right)\,dx=
\frac{\Gamma\left(\nu +1\right)}{\lambda^{\nu+1}}
{_2F}_1\left(\alpha; \nu +1; \gamma \frac{k}{\lambda}\right)dx,
\end{eqnarray*}
and taking account of the relation
\begin{eqnarray*}
{_2F}_1 \left(a, b; c; 1\right) = \frac{\Gamma(c)\Gamma(c-a-b)}
{\Gamma(c-a)\Gamma(c-b)}
\end{eqnarray*}
we derive
\begin{eqnarray}
K^{NN_1}_{jms} = \frac{(\frac{N}{2}-m_+)!
\Gamma(2j+\delta_{1}+\delta_{2}+2)\Gamma\left(j+m_+
+\delta_{1}+\delta_{2}+1\right)}{(j-m_+)!\Gamma\left(\frac{N}{2}+j
+\delta_{1}+\delta_{2}+1\right)}\times \nonumber \\[3mm]
\times{_3F}_2 \left\{
\begin{array}{l}
-N_1,  -j+m_+, j+m_+
+\delta_{1}+\delta_{2}+1 \\
m_1+1,   -\frac{N}{2}+m_+ +1
\end{array}
\biggr| 1 \right\}.  \label{K1}
\end{eqnarray}
The introduction of (\ref{E}) and (\ref{K1}) into (\ref{inter1}) gives
\begin{eqnarray} W^{jms}_{N_{1}N_{2}M_1M_2}
\left(\delta_{1}, \delta_{2}\right) =
\sqrt{\frac{\left(2j+\delta_{1}+\delta_{2}+1\right)\Gamma(N_1+m_1+1)
\Gamma(N_2+m_2+1)}{(N_1)!(N_2)!\left(\frac{N}{2}-j\right)!
(j-m_+)!\Gamma\left(j+m_-
+\delta_{2}+1\right)}}\times \nonumber\\[2mm]
\times \frac{\left(\frac{N}{2}-m_+\right)!}{\Gamma(m_1
+1)}\sqrt{\frac{\Gamma\left(j-m_-
+\delta_{1}+1\right)\Gamma\left(j+m_+
+\delta_{1}+\delta_{2}+1\right)}{\Gamma\left(\frac{N}{2}+j
+\delta_{1}+\delta_{2}+1\right)}}\times  \nonumber \\[2mm]
\times{_3F}_2 \left\{
\begin{array}{l}
-n_1,  -j+m_+, j+m_+
+\delta_{1}+\delta_{2}+1 \\
m_1+1,   -\frac{N}{2}+m_+ +1
\end{array}
\biggr| 1 \right\}\,\delta_{M_1,m+s}\,\delta_{M_2,m-s}.
\label{inter2}
\end{eqnarray}

It is known that the Clebsch-Gordan coefficients for the group $SU(2)$
can be written as \cite{KMP}
\begin{eqnarray}
C_{a \alpha ; b \beta}^{c\gamma} = \left[
\frac{(2c+1)(b-a+c)!(a+\alpha)!(b+\beta)!(c+\gamma)!}
{(b-\beta)!(c-\gamma) ! (a+b-c)!(a-b+c)!(a+b+c+1)!} \right] ^{1/2}
\times \nonumber \\ [3mm] \times
\delta_{\gamma,\alpha+\beta}\frac{(-1)^{a-\alpha}}{\sqrt{(a-\alpha)!}}
\frac{(a+b-\gamma)!}{(b-a+\gamma)!} {_3F}_2 \left\{
\begin{array}{l}
-a+\alpha, c+\gamma+1, -c+\gamma \\
\gamma-a-b,  b-a+\gamma+1  \\
\end{array}
\biggr| 1 \right\}.
\label{CG2}
\end{eqnarray}
Finally, comparing (\ref{inter2}) and (\ref{CG2}), we arrive at the
following representation:
\begin{eqnarray}
W^{jms}_{N_1N_2M_1M_2}\left(\delta_1, \delta_2\right) =
(-1)^{N_1+\frac{m-s+|m-s|}{2}}\,
\delta_{M_1,m+s}\,\delta_{M_2,m-s}\,\times\nonumber \\
\label{W} \\
\times\,C^{j+\frac{\delta_{1}+
\delta_{2}}{2},\,\frac{m_1+m_2}{2}}_{\frac{N+2m_-
+2\delta_2-2}{4},\,\frac{m_2+N_2-N_1}{2};\,\frac{N-2m_-
+2\delta_1-2}{4},\,\frac{m_1+N_1-N_2}{2}}.
\nonumber
\end{eqnarray}
Equation (\ref{W}) proves that the coefficients for the expansion
of the parabolic basis in terms of the spherical basis are nothing
but the analytical continuation, for real values of their
arguments, of the $SU(2)$ Clebsch-Gordan coefficients.

The inverse representation has the form
\begin{eqnarray}
\psi_{Njms}= \sum_{N_1,M_1M_2}\,\tilde{W}^{N_1}_{Njms}\left(\delta_{1},
\delta_{2}\right)\,\psi_{N_1N_2M_1M_2}.
\label{inv-inter}
\end{eqnarray}
The expansion coefficients in (\ref{inv-inter}) are given by the
expression
\begin{eqnarray} \tilde{W}^{N_1}_{Njms}\left(\delta_{1}, \delta_{2}\right)=
(-1)^{N_1+\frac{m-s+|m-s|}{2}}\,
\delta_{M_1,m+s}\,\delta_{M_2,m-s}\,\times\nonumber \\
\label{inv-W} \\
\times\,C^{j+\frac{\delta_{1}+
\delta_{2}}{2},\,\frac{m_1+m_2}{2}}_{\frac{N+2m_-
+2\delta_2-2}{4},\,\frac{N+2m_-+2\delta_2-2}{4}-n_1;\,\frac{N-2m_-
+2\delta_1-2}{4},\,n_1+|m-s|-\frac{N-2m_- -2\delta_1-2}{4}}.
\end{eqnarray}

\section{Prolate Spheroidal Basis}

Let us determine the four-dimensional spheroidal coordinates
\begin{eqnarray}
u_0+iu_1=\frac{d}{2}\,\sqrt{(\xi+1)(1+\eta)}\,e^{i\frac{\alpha+\gamma}{2}},
\qquad
u_2+iu_=\frac{d}{2}\,\sqrt{(\xi-1)(1-\eta)}\,e^{i\frac{\alpha-\gamma}{2}},
\label{scoo}
\end{eqnarray}
where $\xi \in [1,\infty)$, $\eta \in [-1,1]$, and $d$ is the
interfocus distance.

In the spheroidal system of coordinates the four-dimensional double
singular oscillator potential has the form
\begin{eqnarray*}
V = \frac{\mu d^2\omega^2}{2}\left(\xi +\eta\right) +
\frac{4}{d^2}\left[\frac{c_1}{(\xi+1)(1+\eta)}+
\frac{c_2}{(\xi-1)(1-\eta)}\right].
\end{eqnarray*}
In the coordinates (\ref{scoo}) Laplace operator have the form
\begin{eqnarray*}
\frac{\partial^2}{\partial u_i^2} =
\frac{8}{d^2(\xi-\eta)}\Biggl\{\frac{\partial}{\partial \xi}
\left[\left(\xi^2-1\right)\frac{\partial}{\partial \xi}\right]
+ \frac{\partial}{\partial \eta}
\left[\left(1-\eta^2\right)\frac{\partial}{\partial \eta}\right]-
\\ [2mm]
-\frac{1}{2}\left(\frac{1}{\xi+1}-\frac{1}{1+\eta}\right)
\left(\frac{\partial}{\partial \alpha}+
\frac{\partial}{\partial \gamma}\right)^2+
\frac{1}{2}\left(\frac{1}{\xi-1}+\frac{1}{1-\eta}\right)
\left(\frac{\partial}{\partial \alpha}
\frac{\partial}{\partial \gamma}\right)^2\Biggr\}.
\end{eqnarray*}

After the substitution
\begin{eqnarray*}
\psi(\xi,\eta,\alpha,\gamma) = \psi_1(\xi)
\psi_2(\eta)\,e^{im\alpha}\,e^{is\gamma} \end{eqnarray*}
the variables in the Schr\"{o}dinger equation (\ref{schr-so}) are
separated
\begin{eqnarray}
\left[\frac{d}{d\xi}\left(\xi^2-1\right)\frac{d}{d\xi}+\frac{m_1^2}{2(\xi+1)}
-\frac{m_2^2}{2(\xi-1)}-\frac{a^4d^4}{16}\left(\xi^2-1\right)
+\frac{\mu Ed^2}{4\hbar^2}\xi\right]\psi_1 = Q\psi_1,
\label{seq1} \\ [2mm]
\left[\frac{d}{d\eta}\left(1-\eta^2\right)\frac{d}{d\eta}-
\frac{m_1^2}{2(1+\eta)}-\frac{m_2^2}{2(1-\eta)}-
\frac{a^4d^4}{16}\left(1-\eta^2\right)
-\frac{\mu Ed^2}{4\hbar^2}\eta\right]\psi_2 = -Q\psi_2,
\label{seq2} \end{eqnarray}
where $Q$ is a separation constant in spheroidal coordinates.
By eliminating the the energy $E$ from Eqs.(\ref{seq1}) and
(\ref{seq2}), we produce the operator
\begin{eqnarray}
{\hat Q} &=& -\frac{1}{\xi-\eta}
\left\{\eta\frac{\partial}{\partial\xi}\left[\left(\xi^2-1\right)
\frac{\partial}{\partial \xi}\right] +\xi\frac{\partial}{\partial\eta}
\left[\left(1-\eta^2\right)\frac{\partial}{\partial \eta}\right]\right\}-
\nonumber\\ [3mm]
&-& \frac{\xi+\eta+1}{2\left(\xi+1\right)\left(1+\eta\right)}
\left(\frac{\partial}{\partial \alpha}+
\frac{\partial}{\partial \gamma}\right)^2-
\frac{\xi+\eta-1}{2\left(\xi-1\right)\left(1-\eta\right)}
\left(\frac{\partial}{\partial \alpha}-
\frac{\partial}{\partial \gamma}\right)^2+    \label{Q}  \\ [3mm]
&+&\frac{a^4d^4}{16}\left(\xi\eta+1\right)
+\frac{c_1(\xi-\eta)}{2(\xi+1)(1+\eta)}
+\frac{c_2(\xi-\eta)}{2(\xi-1)(1-\eta)}, \nonumber
\end{eqnarray}
the eigenvalues of which are $Q$ and the eigenfunctions
of which are $\psi(\xi,\eta,\alpha,\gamma)$. The significance of the
self-adjoint operator ${\hat Q}$ can be found by switching to
Cartesian coordinates. Passing to the Cartesian coordinates in
(\ref{Q}) and taking (\ref{Lambda2}) and (\ref{Omega2}) into
account, we obtain
\begin{eqnarray} {\hat Q} = {\hat \Lambda} + \frac{a^2d^2}{4}
{\hat \Omega}.
\label{Q1}
\end{eqnarray}
Therefore,
\begin{eqnarray}
{\hat Q}\psi_{Nqms}(\xi,\eta,\alpha,\gamma; d,\delta_1,\delta_2) =
Q_q\psi_{Nqms}(\xi,\eta,\alpha,\gamma; d,\delta_1,\delta_2),
\label{spectr3} \end{eqnarray}
where index $q$ labels the eigenvalues of the operator ${\hat Q}$.

Now construct the spheroidal basis of the four-dimensional double
singular oscillator using the following expressions:
\begin{eqnarray}
\psi_{Nqms} =
\sum_{j=m_+}^{N/2}\,U^{j}_{Nqms}\left(R;\delta_{1},
\delta_{2}\right)\,
\psi_{Njms}.
\label{sinter1}
\end{eqnarray}
\begin{eqnarray}
\psi_{Nqms} =
\sum_{N_1=0}^{\left(N-|M_1|-|M_2|\right)/2}\,
V^{N_1}_{Nqms}\left(R;\delta_{1},
\delta_{2}\right)\,
\psi_{N_1N_2M_1M_2}.
\label{sinter2} \end{eqnarray}
Substituting (\ref{sinter1}) and (\ref{sinter2}) into (\ref{spectr3}),
and then using (\ref{Q1}) we arrive at the following algebraic
equations:
\begin{eqnarray}
\left[Q_{q}-\left(j+\frac{\delta_1+\delta_2}{2}\right)
\left(j+\frac{\delta_1+\delta_2}{2}+1\right)\right]U^{j}_{Nqms} =
\frac{a^2d^2}{4}R\sum_{j'}U^{j'}_{nqms}({\hat \Omega})_{jj'},
\nonumber \\
\label{inter11} \\
\left[Q_{q}-\frac{a^2d^2}{4}
\left(N_1-N_2+\frac{m_1-m_2}{2}\right)\right]V^{N_1}_{Nqms} =
\sum_{N_1'}\,V^{N_1'}_{Nqms}({\hat \Lambda})_{N_1N_1'},
\nonumber \end{eqnarray}
where
\begin{eqnarray*}
({\hat \Omega})_{jj'} = \int\,
\psi_{Njms}^{*}\,{\hat \Omega}\,\psi_{Njms}dV, \qquad
                           = \int\,
\psi_{N_1N_2M_1M_2}^{*}\,{\hat \Lambda}\,\psi_{N_1'N_2'M_1M_2}\,dV.
\end{eqnarray*}
Now using expansions (\ref{W}), (\ref{inv-W}) and formulae \cite{var}
\begin{eqnarray*}
C_{a\alpha;b\beta}^{c\gamma} = -\left[
\frac{4c^2(2c+1)(2c-1)}{(c+\gamma)(c-\gamma)(b-a+c)(a-b+c)
(a+b-c+1)(a+b+c+1)} \right]^{1/2} \times \nonumber \\ [3mm] \times
\Biggl\{\left[\frac{(c-\gamma-1)(c+\gamma-1)(b-a+c-1)(a-b+c-1)(a+b-c+2)
(a+b+c)}{4(c-1)^2 (2c-3)(2c-1)} \right]^{1/2} \times \nonumber\\
[3mm] \times C_{a\alpha;b\beta}^{c-2,\gamma}
-\frac{(\alpha-\beta)c(c-1) - \gamma a(a+1) + \gamma b(b+1)}
{2c(c-1)} C_{a\alpha;b\beta}^{c-1,\gamma}\Biggr\}, \nonumber
\end{eqnarray*}
\begin{eqnarray*}
\left[\,c(c+1) - a(a+1) - b(b+1) - 2\alpha\beta\right]\,
C^{c,\gamma}_{a, \alpha; b, \beta} =  \\ [2mm] =
\sqrt{(a+\alpha)(a-\alpha+1)(b-\beta)(b+\beta+1)}\,
C^{c,\gamma}_{a, \alpha - 1; b, \beta + 1} +
\\ [2mm]
+ \sqrt{(a-\alpha)(a+\alpha+1)(b+\beta)(b-\beta+1)}\,
C^{c,\gamma}_{a, \alpha + 1; b, \beta - 1},
\end{eqnarray*}
and with the orthonormalization conditions
\begin{eqnarray*}
\sum_{\alpha+\beta=\gamma} C_{a\alpha;b\beta}^{c \gamma}
C_{a\alpha;b\beta}^{c'\gamma'} = \delta_{c'c} \delta_{\gamma'\gamma},
\qquad
\sum_{c=|\gamma|}^{a+b} C_{a\alpha ;b\beta }^{c\gamma}
C_{a\alpha';b\beta'}^{c\gamma} = \delta_{\alpha\alpha'}
\delta_{\beta\beta'}
\end{eqnarray*}
for the Clebsch-Gordan coefficients of the group $SU(2)$, for the
matrix elements $({\hat \Omega})_{jj'}$ and $({\hat \Lambda})_{N_1N_1'}$
we get the expressions
\begin{eqnarray}
({\hat \Omega})_{jj'} =
\frac{(m_1+m_2)(m_1-m_2)}{(2j+\delta_{1}+\delta_{2})
(2j+\delta_{1}+\delta_{2}+2)}
\delta_{j',j}-\frac{2
\left(A^{j+1}_{nm}\delta_{j',j+1}+A^{j}_{nm}\delta_{j',j-1}\right)}
{2j+\delta_{1}+\delta_{2}},
\label{inter16}
\end{eqnarray}
\begin{eqnarray}
({\hat \Lambda})_{N_1N_1'} = \biggl[(N_1+1)(N_2+m_-) +
(\frac{N}{2}-N_1+\delta_2)(N_1+|m-s|+\delta_2)
+ \nonumber \\[2mm]
+m_-(m_+
+\delta_2)\frac{1}{4}(\delta_1-\delta_2)(\delta_1-\delta_2-2)
 +\biggr]\, \delta_{N_1'N_1} -\nonumber \\
\label{inter25}\\
-\sqrt{n_2(N_1+1)(N_1+m_1+1)
(N_2+m_2)}\delta_{N_1',N_1+1}-\nonumber \\[2mm]
- \sqrt{N_1(N_2+1)(N_1+m_1+1)
(N_2+m_2+1)}\delta_{N_1',N_1-1}  \nonumber
\end{eqnarray}
where
\begin{eqnarray*}
A^{j}_{Nms} &=& \sqrt{(j-m_- +\delta_1)(j+m_- +\delta_2)}\times \\ [3mm]
&\times&
\left[\frac{(j-m_+)(j+m_+ +\delta_{1}+\delta_{2})
(N-2j)(N+2j+2\delta_{1}+2\delta_{2})}
{4\left(j+\frac{\delta_{1}+\delta_{2}}{2}\right)^2
(2j+\delta_{1}+\delta_{2}-1)(2j+\delta_{1}+\delta_{2}+1)}\right]^{1/2}.
\end{eqnarray*}
 Substituting expressions (\ref{inter16}) and (\ref{inter25}) into the
algebraic equations (\ref{inter11}), we derive the three-term recursion
relations
\begin{eqnarray*}
A^{j+1}_{Nms}U^{j+1}_{Nqms}+A^{j}_{nms}U^{j-1}_{Nnqms}&=&
\Biggl\{\frac{4}{a^2d^2}\left[Q-\left(j+\frac{\delta_1+\delta_2}{2}\right)
\left(j+\frac{\delta_1+\delta_2}{2}+1\right)\right]- \\ [3mm]
&-&\frac{(m_1+m_2)(m_1-m_2)}{(2j+\delta_{1}+\delta_{2})
(2j+\delta_{1}+\delta_{2}+2)}
\Biggr\}U^{j}_{Nqms}
\end{eqnarray*}
\begin{eqnarray*}
\biggl[(N_1+1)(N_2+m_-) +
(N-N_1+\delta_2)(N_1+m_2)
+\frac{1}{4}(\delta_1-\delta_2)(\delta_1-\delta_2-2)
 + \nonumber \\[2mm]
+m_-(m_+ +\delta_2)+\frac{a^2d^2}{2}
\left(N_1-N_2+\frac{m_1-m_2}{2}\right)-Q_{q}\biggr]\,
V^{N_1}_{Nqms}= \nonumber \\ [2mm]
=\sqrt{N_2(N_1+1)(N_1+m_1+1)
(N_2+m_2)}\,V^{N_1+1}_{Nqms}+ \nonumber \\[2mm]
+\sqrt{N_1(N_2+1)(N_1+m_1+1)
(N_2+m_2+1)}\,V^{N_1-1}_{Nqms}
\end{eqnarray*}
for the interbasis expansions coefficients $U^{j}_{Nqms}$ and
$V^{N_1}_{Nqms}$.

\vspace{5mm}

{\bf Acknowledgements.}
The author is grateful to Dr. L.G.~Mardoyan for many
helpful dicussions.

\end{document}